\documentclass[cha,aip,amsmath,amssymb,preprint,reprint]{revtex4-1}
\usepackage{graphicx}
\usepackage{dcolumn}
\usepackage{bm}
\usepackage{multirow}
\usepackage{mathptmx}
\usepackage[utf8]{inputenc}
\usepackage[T1]{fontenc}
\usepackage{etoolbox}
\usepackage{xcolor}
\usepackage[colorlinks=true,allcolors=blue]{hyperref}
\usepackage{hyperref}
\usepackage{cleveref}

\makeatletter
\def\@email#1#2{%
 \endgroup
 \patchcmd{\titleblock@produce}
  {\frontmatter@RRAPformat}
{\frontmatter@RRAPformat{\produce@RRAP{*#1\href{mailto:#2}{#2}}}\frontmatter@RRAPformat}
  {}{}
}%
\makeatother
\begin{document}

\preprint{AIP/123-QED}

\title{Bright, directional electron emission from nanowire coated targets under petawatt, femtosecond irradiation}

\author{Ameya Parab}

 \thanks{Email: parabameya102@gmail.com}
 \affiliation{Tata Institute of Fundamental Research, Colaba, Mumbai 400005, India}

\author{Jian Fuh Ong}%
    \thanks{Email: jianfuh.ong@eli-np.ro}
\affiliation{ 
Extreme Light Infrastructure - Nuclear Physics (ELI-NP), ``Horia Hulubei'' National Institute for R\&D in Physics and Nuclear Engineering (IFIN-HH), 30 Reactorului Street, Bucharest-M\u{a}gurele, 077125, Romania}

\author{Stefania Ionescu}
\affiliation{ 
Extreme Light Infrastructure - Nuclear Physics (ELI-NP), ``Horia Hulubei'' National Institute for R\&D in Physics and Nuclear Engineering (IFIN-HH), 30 Reactorului Street, Bucharest-M\u{a}gurele, 077125, Romania}
\affiliation{National University of Science and Technology POLITEHNICA Bucharest, Splaiul Independentei no. 313, Bucharest, Romania}
 
\author{Sagar Dam}
\affiliation{Tata Institute of Fundamental Research, Colaba, Mumbai 400005, India}

\author{Sk Rakeeb}
\affiliation{Tata Institute of Fundamental Research, Colaba, Mumbai 400005, India}

\author{Hideaki Habara}
\affiliation{University of Osaka, Yamadaoka, Suita, Osaka 565-0871, Japan}

\author{Y.Keita}
\affiliation{University of Osaka, Yamadaoka, Suita, Osaka 565-0871, Japan}

\author{Rudrajyoti Palit}
 \affiliation{Tata Institute of Fundamental Research, Colaba, Mumbai 400005, India}

\author{Daniel Popa}
\affiliation{ 
Extreme Light Infrastructure - Nuclear Physics (ELI-NP), ``Horia Hulubei'' National Institute for R\&D in Physics and Nuclear Engineering (IFIN-HH), 30 Reactorului Street, Bucharest-M\u{a}gurele, 077125, Romania}

\author{Deepak Sangwan}
\affiliation{ 
Extreme Light Infrastructure - Nuclear Physics (ELI-NP), ``Horia Hulubei'' National Institute for R\&D in Physics and Nuclear Engineering (IFIN-HH), 30 Reactorului Street, Bucharest-M\u{a}gurele, 077125, Romania}

\author{Klaus Spohr}
\affiliation{ 
Extreme Light Infrastructure - Nuclear Physics (ELI-NP), ``Horia Hulubei'' National Institute for R\&D in Physics and Nuclear Engineering (IFIN-HH), 30 Reactorului Street, Bucharest-M\u{a}gurele, 077125, Romania}

\author{Lucian Tudor}
\affiliation{ 
Extreme Light Infrastructure - Nuclear Physics (ELI-NP), ``Horia Hulubei'' National Institute for R\&D in Physics and Nuclear Engineering (IFIN-HH), 30 Reactorului Street, Bucharest-M\u{a}gurele, 077125, Romania}

\author{Adrian Vatcu}
\affiliation{ 
Extreme Light Infrastructure - Nuclear Physics (ELI-NP), ``Horia Hulubei'' National Institute for R\&D in Physics and Nuclear Engineering (IFIN-HH), 30 Reactorului Street, Bucharest-M\u{a}gurele, 077125, Romania}

\author{Prashant Kumar Singh}
\affiliation{Tata Institute of Fundamental Research, Gopanapally, Serilingampally, Telangana 500046, India.}

\author{Kazuo A. Tanaka}
\affiliation{ 
Extreme Light Infrastructure - Nuclear Physics (ELI-NP), ``Horia Hulubei'' National Institute for R\&D in Physics and Nuclear Engineering (IFIN-HH), 30 Reactorului Street, Bucharest-M\u{a}gurele, 077125, Romania}
\affiliation{University of Osaka, Yamadaoka, Suita, Osaka 565-0871, Japan}

\author{G.Ravindra Kumar}
    \thanks{Corresponding author: grk@tifr.res.in}
    \affiliation{Tata Institute of Fundamental Research, Colaba, Mumbai 400005, India}

\date{\today}

\begin{abstract}
Interactions of relativistically intense laser pulses with structured targets have long been explored for controlling energy absorption and particle acceleration. However, at upcoming multi-petawatt laser facilities, the survivability of such nanostructures under realistic temporal contrast conditions remains a key concern. We report an experimental and simulation study of nanowire targets irradiated by the ELI-NP 1-PW laser without a plasma mirror. At the built in, readily available contrast of $10^{-10}$, the nanowires survive the laser pre-pulse and produce a robust enhancement in relativistic electron flux, energy, and directional emission compared to flat targets indicating that at better contrasts they can show similar enhancement at the 10 PW level. These results establish nanowire targets as resilient and reliable tools for relativistic electron manipulation at state of the art facilities.

\end{abstract}

\maketitle
   
Relativistic electron generation is a central process in ultra-intense laser–solid interactions\cite{Rocca:24,mourou_Tajima_2006optics_RMP,fortov_Extreme_states_of_matter,Kruer,PGibbon_1996,10.1063/1.860697}, governing energy transport in the target, emission of secondary radiation, and formation of extreme electromagnetic fields. Control over the production and dynamics of these electrons is therefore crucial for both fundamental studies as well as applications ranging from high-energy-density physics to secondary particle and radiation sources \cite{moreau2019enhanced}.

In parallel, nanophotonics has emerged over the past decade as a mature field focused on the manipulation of optical fields through sub-wavelength structuring of matter \cite{ciappina2017attosecond,prl_jiang2016microengineering}. By tailoring geometry at the nanoscale, nanophotonic systems enable precise control over key optical properties such as reflectivity, absorption, local field enhancement, and polarization \cite{nat_photonics_purvis2013relativistic,prl_kaymak2016nanoscale,nat_photonics_hollinger2020extreme,pop_eftekhari2022laser,scientific_reports_fedeli2018ultra,PP_rajeev_PRL_2003,S_Kahalay_PRL_2008}. When nanophotonic concepts are combined with ultra-intense and ultrashort laser–matter interactions, they give rise to the emerging field of plasma nanophotonics \cite{nat_photonics_purvis2013relativistic}. In this regime, nanostructured targets dynamically evolve into dense plasmas on femtosecond timescales, while simultaneously shaping the incident laser fields and the ensuing plasma response. Despite its promise, plasma nanophotonics remains largely unexplored, particularly in the relativistic intensity regime, where complex interplay between laser fields, evolving plasma structures, and electron acceleration mechanisms can lead to qualitatively new behavior. 
\begin{figure*}[!ht]
   \centering
   \includegraphics[width=1\linewidth]{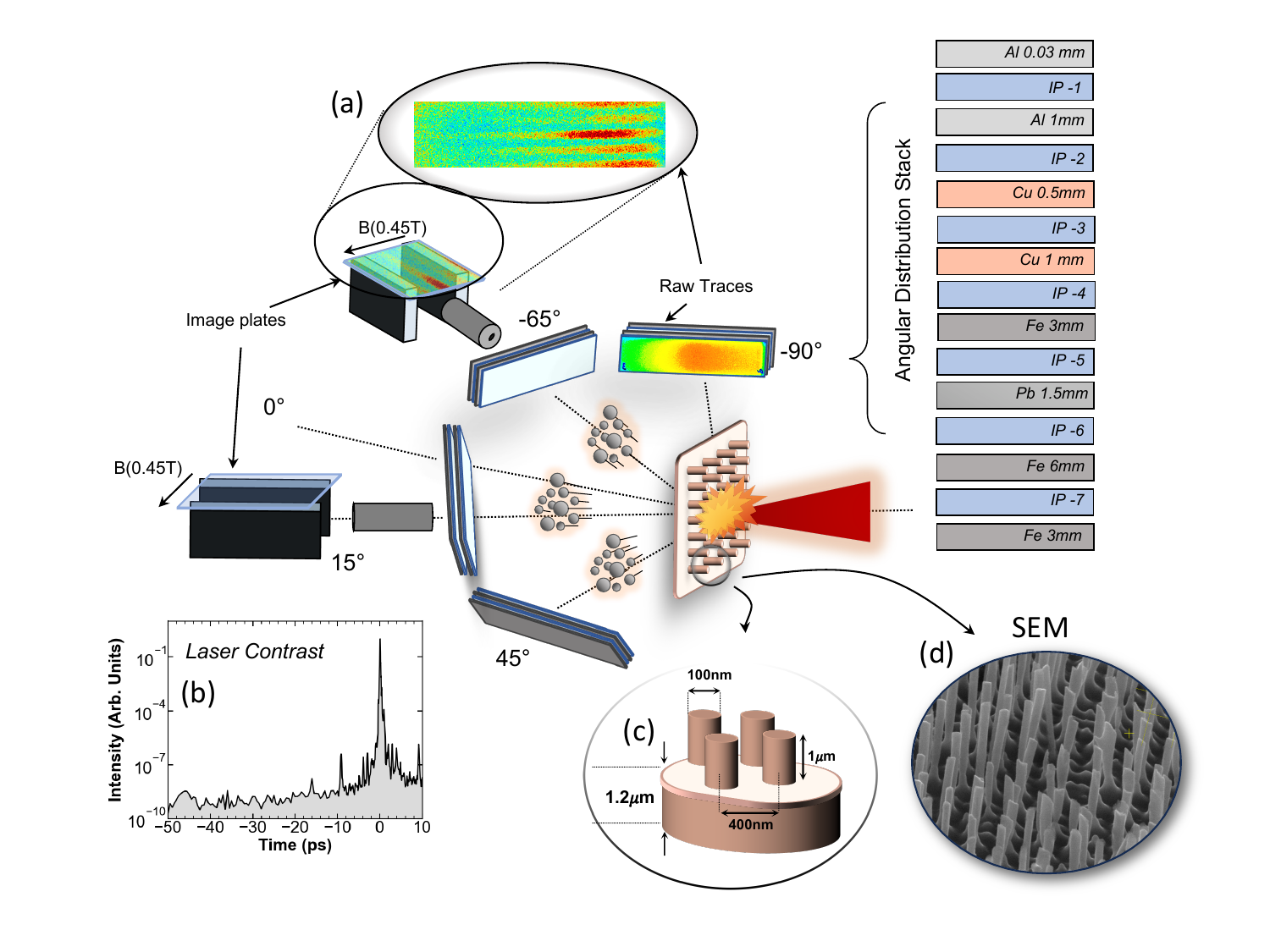}
   \caption{Figure 1. (a) Schematic of the experimental setup. The top-right panel shows the each of the four detector stacks used to measure the angular distribution of emitted electrons; (b) Laser temporal contrast measurement showing a contrast of $\sim10^{-10}$ approximately 50 ps before the main pulse. (c) Schematic of the nanowire target geometry and structural parameters. (d) Scanning electron microscope (SEM) image of the fabricated nanowire array.}
   \label{fig:1:setup}
\end{figure*}

Exploring this regime offers a pathway to uncover novel electron generation mechanisms and to achieve unprecedented control over relativistic laser–plasma interactions \cite{pop_cao2010enhanced,PRR_aprajit_2025,Sci-rep_amit_lad}. The charged particles emitted during such laser–plasma interactions are strongly influenced by the instantaneous electromagnetic field configuration at the interaction site. These field structures can be tailored through coherent field control \cite{prr_ong2021electron,pop_park2021absolute,apl_ji2010efficient,laserphotonics_ankit,pop_zhao2010acceleration}, often achieved by employing corrugated or nanostructures targets. Coupling intense laser pulses to such engineered nanostructures has demonstrated remarkable potential in controlling the generation and emission characteristics of particles in the plasma \cite{electron_nanowire_habara,parameter_scan_elec_temp,bargsten2017energy,prx_samsonova2019relativistic}. 

In this work, we present a combined experimental and simulation study of relativistic electron emission from nanowire targets at the ELI-NP facility\cite{commissioning_1PW,commissoning_1PW_2}. The experiments were performed at peak laser intensities of $\sim 3 \times 10^{21}\,\mathrm{W/cm^{-2}}$ $(a_0=36)$ with an inbuilt temporal contrast of $\sim 10^{-10}$ 50 ps before the main pulse. Using energy-resolved image-plate stacks, electron spectrometers, and angularly resolved diagnostics, we observe a clear enhancement of electron yield from nanowire targets compared to flat foils. More importantly, nanowires exhibit pronounced energy-dependent anisotropy in the emitted electron distribution, a feature that is absent in flat targets under otherwise identical conditions. 3D Particle-In-Cell simulations reproduce these experimental trends and reveal that the observed angular structures persist over a finite range of pre-plasma scale lengths. This demonstrates that nanostructure-induced control of relativistic electron phase space remains effective even in the absence of ultra-high temporal contrast \cite{mourou_Tajima_2006optics_RMP,kiriyama2012temporal,choi2020highly,contrast1998itatani,plasmamirror2006wittmanntowards,plasmamirror2018foldes,preplasma_nanowire,ankit2022spectralinterferometry}. These results establish nanowire targets as robust phase-space modifiers of relativistic electron emission and provide new insight into electron acceleration and transport in laser–solid interactions under experimentally relevant conditions.

\begin{figure*}
   \centering
   \includegraphics[width=\linewidth]{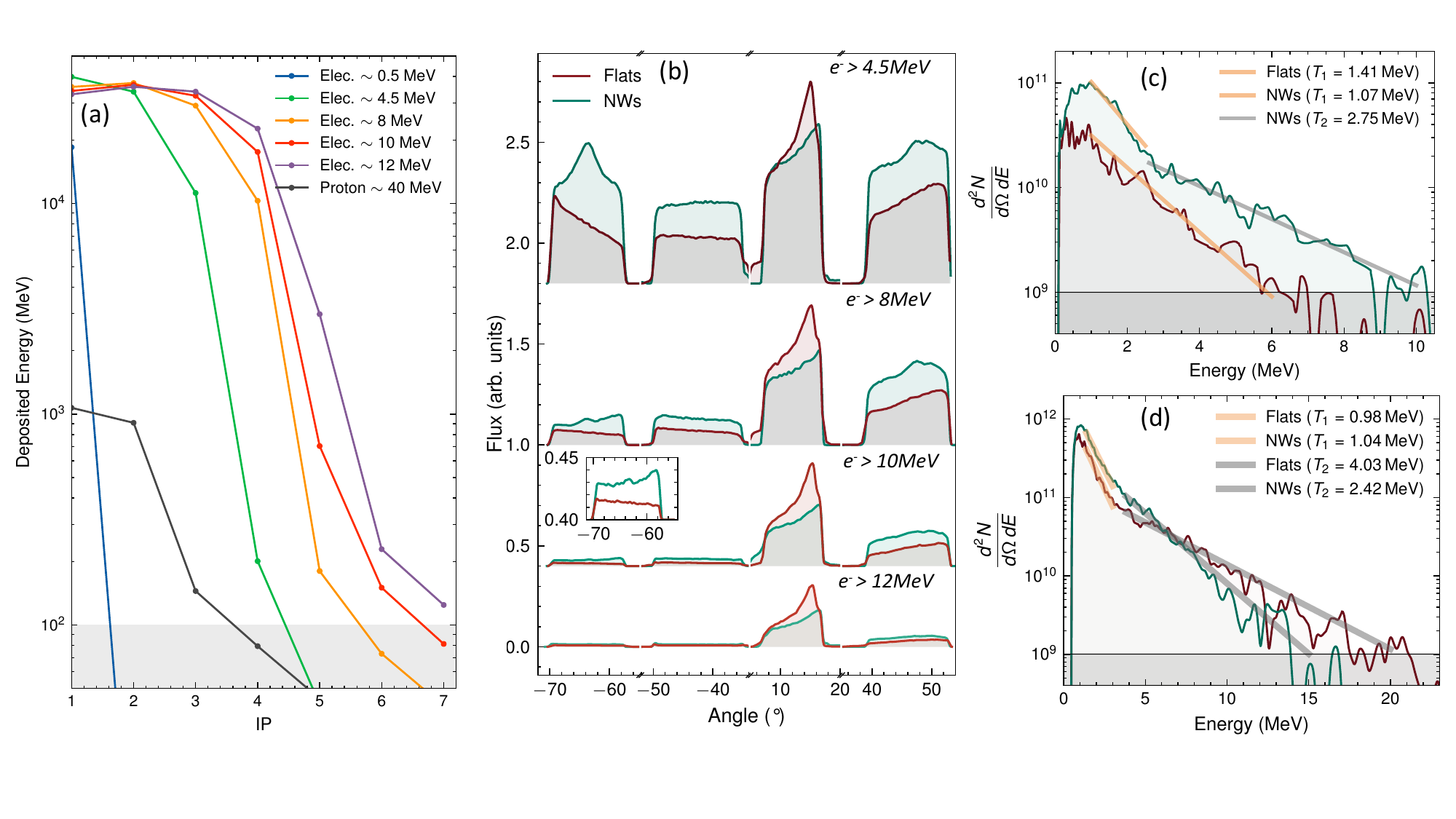}
   \caption{ (a) GEANT4 simulations of the detector stack response to incident electrons of varying energies and proton beams, used to estimate the contribution of proton contamination in the signal recorded by the first imaging plate (IP) layers.
(b) Energy-resolved angular distribution of the emitted electrons, displayed in ascending order from top to bottom.
(c,d) Electron energy spectra measured along the $-64^\circ$ and $15^\circ$ directions, respectively.}
   \label{fig:geant&ang_dist}
\end{figure*}

The experiments were carried out at the ELI-NP facility in Romania, using a high-power laser delivering approximately 15 J of energy onto the target. A 19 cm diameter laser beam was focused onto the target using an f/3.7 off-axis parabolic mirror, achieving a focal spot size of approximately $4.4\ \mu\text{m}$. The laser pulse duration at focus was 23 fs, leading to a peak intensity of about $3 \times 10^{21}\ \text{W/cm}^2$, corresponding to a normalized vector potential of $a_0 \approx 36$—well into the super relativistic regime. The targets irradiated during the experiment were surface coated with Nickel nanowires as well as nickel flats the latter as a reference. The parameters used for the wires are shown in Fig. \ref{fig:1:setup} (inset) with Scanning Electron Microscope (SEM) image. The detail fabrication is discussed in the Methods section.

All diagnostics measured signals from the rear side of the target. Angularly resolved measurements are a key diagnostic for characterizing the directionality and divergence of relativistic electrons emitted from laser–solid interactions. Based on the angular distributions, specific angles of interest were identified, and energy spectra at these angles were subsequently analyzed to gain deeper insight into the dominant electron acceleration mechanisms.
At the high laser intensities explored in this work, commonly used detectors such as image plates (IPs) are sensitive not only to electrons but also to bremsstrahlung background and ion-induced signals. Suppressing these unwanted contributions is therefore essential for reliable electron measurements. To achieve this, multiple IPs were arranged in stacks separated by metallic filters of varying atomic number, as previously reported in literature \cite{Wong2024}. Seven such IP stacks were employed to obtain energy-resolved angular distributions of electrons. The stacks were positioned to maximize angular coverage within the constraints of the experimental geometry.

The filter sequence was designed with low-Z materials at the front, followed by progressively higher-Z filters deeper in the stack. Consequently, the front IP's are sensitive to lower-energy particles, while deeper IP's preferentially record higher-energy components. The detailed ordering of the stack is shown schematically in the corresponding figure.

To quantitatively disentangle electron and proton contributions, and to assess the impact of bremsstrahlung generated within the stack, detailed GEANT4 Monte Carlo simulations were performed to model the IP response. These simulations provide a robust framework for interpreting the experimentally measured angular and energy distributions. The details of the Geant4 simulations are discussed in the methods section.

The GEANT4 results of the simulation are presented in Fig. \ref{fig:geant&ang_dist} (a), which shows the total deposited energy of protons and electrons as a function of IP layer number. A clear trend is observed: while both species contribute significantly to the signal in the first three layers, the proton contribution drops sharply beyond IP 3, becoming nearly negligible from IP 4 onward. By contrast, electrons continue to penetrate deeper into the stack, with their relative contribution dominating from IP 4 to IP 7.

Based on these results, the analysis presented in this work focuses exclusively on layers 4–7 of the stack. Within this range, a natural energy discrimination arises: lower-energy electrons deposit most of their energy in the shallower layers (particularly IP 4), whereas higher-energy electrons are more likely to traverse further and deposit predominantly in the deeper layers (e.g. IP 7). Although this approach effectively removes the proton background, it introduces a limitation: by discarding IP 1–3, we lose sensitivity to the lowest-energy electrons, which would have been stopped in the front layers.  The angular-distribution stacks were positioned at four discrete locations, optimized within the experimental constraints to achieve the widest possible angular coverage.

\begin{figure*}
   \centering
   \includegraphics[width=\linewidth]{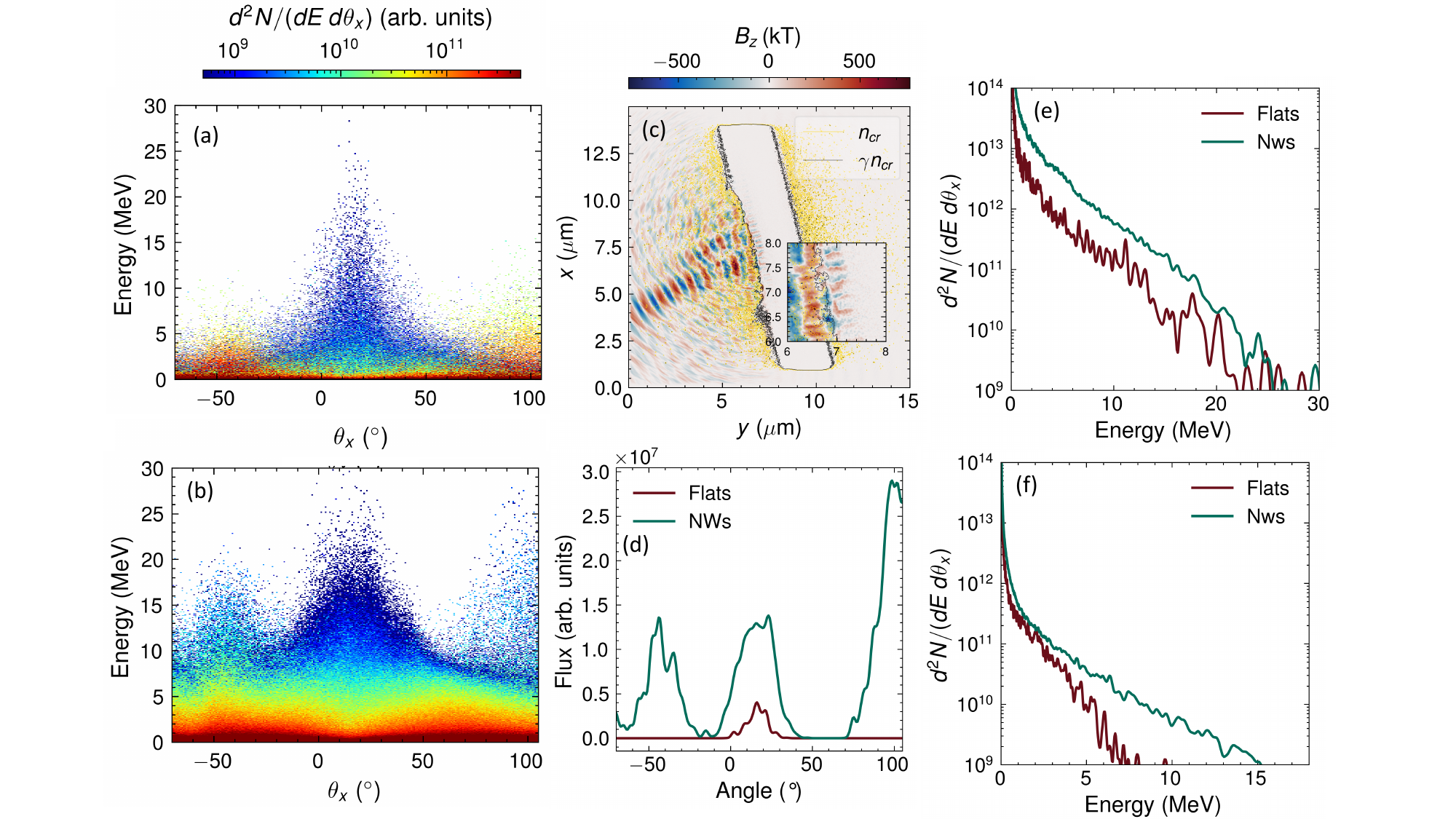}
   \caption{3D PIC simulation results detailing the electron emission. The energy-resolved angular distribution of forward-emitted electron for (a) flat, and (b) nanowire targets. (c) The snapshot of the reflected laser light from the nanowire target. The laser propagates in the $+y$-direction and incident on the target tilted at $15^\circ$. The quasi-static azimuthal magnetic field is generated between the wires as shown in the inset. The yellow and black contour denoted the critical density, $n_{cr}$ and the relativistic critical density, $\gamma n_{cr}$ surface, respectively. (d) The energy-integrated angular distribution for both targets. The energy spectra at a selected direction (e) $\theta_x = 15^\circ$, and (f) $\theta_x = -64^\circ$, integrated within $\pm 10^\circ$ forward cone.
   }
   \label{fig:simulationangdist}
\end{figure*}

Figure~\ref{fig:geant&ang_dist}(b) shows the angular distribution recorded in IP layers~4--7, arranged from top to bottom, covering the angular range from $-71^\circ$ to $55^\circ$. A pronounced feature is observed around $-64^\circ$, where a strong electron flux is detected across all layers in the nanowire targets compared to the nickel flat foils. A secondary maximum is also visible near $45^\circ$ in the nanowires, again exceeding the corresponding signal in the flats.  

The flat targets show an increased flux at $15^\circ$ along with high angular cutoff that becomes more evident in the deeper layers (up to IP~7). Based on these observations, electron spectrometers (ESMs) with a  uniform magnetic field of around 0.45T were positioned at $-64^\circ$ and $15^\circ$ to probe the electron spectra along these characteristic directions. The measured spectra are presented in Figure \ref{fig:geant&ang_dist}: panel~(c) shows the spectrum recorded at $-64^\circ$, while panel~(d) corresponds to $15^\circ$. The electron cutoff energies were extracted from the spectra at the two diagnostic angles. At $-64^\circ$, the cutoff for nanowires was found to be 9.91~MeV, compared to 5.47~MeV for flat targets. At $15^\circ$, the cutoff was higher for both cases: 13.94~MeV for nanowires and 18.71~MeV for flats. Thus, the cutoff energy increases for nanowires at $-64^\circ$, whereas at $15^\circ$ the opposite trend is observed, with flats extending to higher energies.

To further characterize the electron distributions, the spectra were fitted with a double-Maxwellian function corresponding to a two-temperature model. At $-64^\circ$, nanowires exhibited a hot electron temperature of 1.07~MeV and a hotter component of 2.75~MeV in contrast to flats, which were well described by a single slope temperature those for of 1.36~MeV. At $15^\circ$, both targets showed two-temperature signatures: for nanowires, the hot electron temperature was 1.04~MeV and for flats 0.98~MeV, nearly identical. The hotter component, however, reached 2.43~MeV for nanowires and 4.03~MeV for flats.  
The extracted parameters are summarized in Table~\ref{tab:cutoff}
\begin{table}
\centering
\begin{minipage}{0.5\textwidth}
\centering
\caption{Electron cutoff energies and fitted temperatures for nanowire and flat targets at $-64^\circ$ and $15^\circ$.}
\label{tab:cutoff}
\begin{tabular}{lcccc}
\hline
\textbf{Angle} & \textbf{Target} & \textbf{Cutoff} & \textbf{Cold $T_e$} & \textbf{Hot $T_e$} \\
               &                 & (MeV)           & (MeV)               & (MeV) \\
\hline
\multirow{2}{*}{$-64^\circ$} & NWs & 9.91 & 1.07 & 2.75 \\
                             & Flats     & 5.47 & 1.36 & -- \\
\hline
\multirow{2}{*}{$15^\circ$}  & NWs & 13.94 & 1.04 & 2.43 \\
                             & Flats     & 18.71 & 0.98 & 4.03 \\
\hline
\end{tabular}
\end{minipage}
\end{table}

To support the experimental observations, combined three-dimensional radiation hydrodynamic (3D RHD) and particle-in-cell (3D PIC) simulations were carried out. Hydrodynamic simulations were performed to model the effects the laser pre-pulses on nanowire targets. The expanded target profile was then converted to 3D PIC simulations to model the interaction with the main pulse. The simulation parameters were selected to closely replicate the experimental conditions. Details of the numerical parameters and the setup are described in the Methods section.
\begin{figure*}
   \centering
   \includegraphics[width=0.8\linewidth]{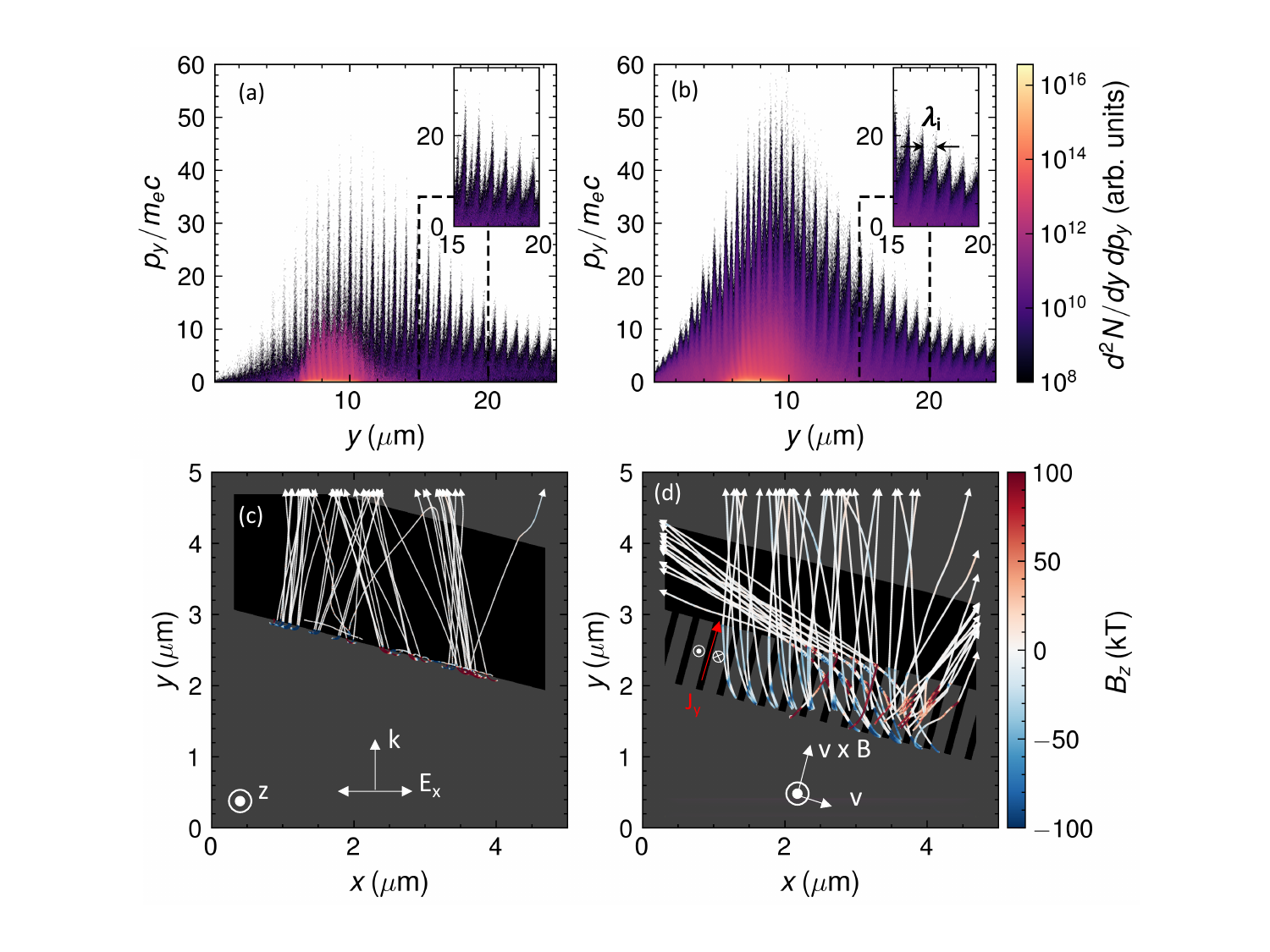}
   \caption{Electrons $y-p_y$ phase space plot for (a) flat, and (b) nanowire target within the angular region of $10^\circ < \theta_x < 20^\circ$. The insets show the enlarged region at the target rear side with peak-to-peak separation of $\lambda$. Electron trajectories from 2D PIC simulation for (c) flat, and (d) nanowire target. The laser is propagating in the $+y-$direction and polarized in $x-$direction. The $z$ axis is directed out of the page. The color along the trajectories is the magnetic field component $B_z$ at electron position. The red arrow in (d) annotates direction of the return current $J_y$. The direction of quasi-static azimuthal magnetic field generated by this return current is shown. Electrons strip off by the laser field is accelerated forward by the $\mathbf{v} \times \mathbf{B}$ force.}
   \label{fig:5}
\end{figure*}
Figures \ref{fig:simulationangdist} (a) and (b) show the energy resolved angular distribution of the forward-emitted electron for flat and nanowire targets, respectively. Both spectra show the highest number of electrons in the laser propagation direction ($\theta_x \approx 15^\circ$) with energies greater than $20 \, \mathrm{MeV}$. The laser is propagating in the $+y-$direction and polarized in the $x-$direction. Figure \ref{fig:simulationangdist} (c) shows the snapshot of laser main pulse reflected off the surface of the relativistic critical density, $\gamma n_{cr}$. At normalized intensity, $a_0 = 36$, the plasma with density $n_e<\gamma n_{cr}$ is transparent to the laser field due to relativistic self-induced transparency, where $\gamma=(1+a_0^2/2)^{1/2}$, and $n_{cr}=\epsilon_0m\omega_\mathrm{L}^2/e^2$ is the critical density. Here, the relativistic critical density is $\gamma n_{cr}=4.43\times 10^{22}\,\mathrm{cm^{-3}}$ and is denoted by the black contour in Fig. \ref{fig:simulationangdist} (c). The void among the nanowires has been filled by plasma as a result of the interaction with the laser pre-pulse and rising edge of the main pulse. As the main pulse approaches, the strong ponderomotive pushes the $\gamma n_{cr}$ surface inwards corroding the nanowire. At the same time, we observe the generation of quasi-static azimuthal magnetic field of the order of $\sim 100 \, \mathrm{kT}$ as shown in the inset of Fig. \ref{fig:simulationangdist} (c) due to the nanoscale z-pinch \cite{prl_kaymak2016nanoscale} despite the laser field being shuttered from penetrating the nanowire. This magnetic field provides additional evidence for the persistence of the nanowire structure at high intensities.

Two pronounced peaks were observed at $60^\circ < \theta_x < 90^\circ$ and $ \theta_x < 40^\circ$ for nanowire, which are notably suppressed for flat targets. These peaks are enhanced not only in terms of energies but also in flux, as shown in Fig. \ref{fig:simulationangdist} (d). The electron flux in the angular range $10^\circ < \theta_x < 20^\circ$ is found to be comparatively lower than that observed in the NWs case. This discrepancy with respect to experimental measurements primarily arises from differences in the collection conditions. In the simulations, all electrons within the computational domain are accounted for, whereas in experiments the detectors are located several meters away from the interaction region, effectively sampling only those electrons that propagate large distances.
As a consequence, in the simulation enhanced absorption and the generation of recirculating electrons, as discussed in the literature, lead to an apparent increase in the simulated flux within this angular range. However, a significant fraction of these recirculating electrons do not reach the detectors in the experimental configuration, thereby resulting in the observed deviation between simulation and experiment. The electron spectra in the direction of $\theta_x = 15^\circ$, and $\theta_x = -64^\circ$ were integrated within a cone angle of $4^\circ$  to reproduce close resemblance to the experimental measurement and are shown in Fig. \ref{fig:simulationangdist} (e) and (f), respectively. The simulation results of electron spectra in these directions show enhancement for nanwoire over the flat target.

The electron distribution at the laser incident direction is attributed to resonance absorption owing to the interaction of p-polarized laser at oblique incident with the target critical density surface. This is evident in Figs. \ref{fig:5} (a) and (b) phase space $y-p_y$ plots, where the peak-to-peak distance of the spectra is separated by distance of $\lambda_L$. To explain the excess of the lateral peaks in nanowire targets, we performed reduced simulations in 2D to track its trajectories. A sample of electron trajectories and the magnetic field at its location are depicted in Fig. \ref{fig:5} (c) and (d) for flat and nanowire targets, respectively. In Fig. \ref{fig:5} (c), electrons stripped off by the $E_x$ component of the laser field during the first-half of the cycle are pushed back into the target at the second-half cycle and travel in one direction. Electrons pulled out during the second cycle travel in the opposite direction. 

For nanowire target shown in Fig. \ref{fig:5} (d), the $E_x$ component of the laser field strips off electrons at the nanowire tip and the $B_z$ component accelerates it forward. A return current is then generated on the nanowire surface, flowing toward the substrate to maintain local current quasi-neutrality. This return current then generates a quasi-static azimuthal magnetic field around the nanowires with magnitudes of the order of hundreds of kilo-Tesla. Electrons traveling close to the nanowire surface will be subjected to further forward acceleration by this field. Electrons are deflected in clear two distinct directions at the end of the nanowires, and the direction is determined by the sign of the magnetic field at the electron position. This explains the excess of lateral peaks in nanowire spectra.

To the best of our knowledge, these results provide the first clear evidence of directional change of electron emission from nanowire targets. A key finding was that the directional change of electron emission in the nanowire target is attributed to the quasistatic magnetic field generated by the return current. This provides an indirect indication of nanoscale z-pinch. In other words, the nature of the laser-nanowire target interaction has redistributed the electron emission direction, which expect to subsequently affect the proton acceleration and gamma-ray emission.

\section{Conclusions}
 We have demonstrated that the readily available contrast  at the ELI-NP facility ($10^{-10}$) is adequate to show case the robustness of the nanostructure coated targets for fast electron enhancement.  This gives confidence that the plasma mirror aided contrast ($10^{-12}$)  promises  even more impressive enhancements at the higher intensities, as predicted by simulations. We also notice three prominent emission peaks observed for the nanowire targets, reminiscent of  diffraction of light from a grating. The collective response of the nanowire array leads to coherent addition of the electron currents, resulting in efficient flux enhancement along these discrete directions.  More investigations are thus necessary to establish the  place of nanostructures  at  ultrarelativistic intensities in the multi petawatt range. 
 
\section*{Methods}
\subsection*{Nanostructure growth}
The nanostructured targets are grown in the Target Laboratory from ELI-NP, using bottom-up methods, involving electrochemical deposition of metals from a Watts bath in a porous template obtained by aluminum anodization in acidic electrolytes \cite{Gheorghiu2021}. The nanowire characteristics are tailored through the process parameters as follows: the diameter and gap between structures is given by the template and therefore, by the anodization electrolyte and the anodization voltage; the length of the nanowires and the thickness of the substrate is modified by the deposition time; and the material is depending on the electrodeposition bath, made of soluble metal salts, which for nickel, used in our experiments contains nickel chloride and nickel sulfate\cite{Ionescu2025}. To prepare the template, a 2 stage anodization process is performed to ensure smooth and parallel walls \cite{Masuda1995}, followed by a pore widening and opening of the pores on both sides. A seed layer is sputtered on one side and after the electrodeposition, the template is removed in sodium hydroxide bath, followed by the drying of the sample in solvents with lower surface tension to reduce the structures clustering. Similarly, nickel flat targets were prepared.

\subsection*{Geant4 Simulations}
In the simulation, a stack of image plates was constructed with the exact material composition and layer thicknesses corresponding to the experimental configuration. The source was modeled as a point-like emitter positioned approximately 100 mm upstream of the first IP layer, consistent with the experimental geometry. A uniform energy distribution of particles was assumed for input: electrons extending up to 30 MeV and protons up to 40 MeV, each initialized with a divergence half-angle of 5°. This configuration allowed us to track the transport, energy loss, and final deposition of both species within the layered IP detector.

\subsection*{Hydrosimulations and PIC Siumaltions}
Three-dimensional radiation hydrodynamic (3D RHD) simulations were performed using the \textsc{flash} code version 4.8 \cite{flash}. The size of the simulation box is $15 \times 15 \times 15 \, \mathrm{\mu m^3}$. We used Adaptive Mesh Refinement with $6$ levels of refinement over the mass density, achieving a minimum resolution of $29 \, \mathrm{nm}$. The initial mass density of nickel is $\rho_0 = 8.9 \, \mathrm{g \, cm^{-3}}$ with an average charge $\overline{Z}=28$, and a mass number $\overline{A}  = 58.7$. The pressure is obtained from tabulated equations of state data \textsc{sesame} and opacity data from \textsc{tops}. The ionization state was calculated using the Thomas-Fermi model. We used the Lee-More model for heat exchange and electron heat conduction, with a flux limiter of $0.08$. Radiation diffusion is incorporated using multi-group diffusion theory with $20$ energy groups.
The flat nickel target has a thickness of $2 \, \mathrm{\mu m}$. The nickel nanowire is $30 \times 30$ aligned array with diameter $d = 100\, \mathrm{nm}$, length $L = 1 \,\mathrm{\mu m}$, and $300 \, \mathrm{nm}$ center-to-center separation attaching on $1.2 \,\mathrm{\mu m}$ substrate of the same material. The laser is incident at $15^\circ$ from the target normal. The laser temporal profile is adopted from the fitted curve in Fig. \ref{fig:1:setup} (d), and sampled with $64$ points in logarithmic scale from the start of ablation up to $t \sim -0.24 \, \mathrm{ps}$. 

For the flat target, the ablation starts from $t \sim -50 \, \mathrm{ps}$, and for the nanowire target, the ablation starts from $t \sim -311 \, \mathrm{ps}$. The estimation of the onset of ablation was carried out by comparing the absorbed fluence to the fluence threshold in the long pulse ablation along the contrast profile. The radius of the focal point is $R_x = R_z = 2.3 \, \mathrm{\mu m}$, where $R_x = d_\mathrm{FWHM} / (2 \sqrt{\ln 2}) = w_0 / \sqrt{2}$. The spatial profile in the focus spot has the form $I(r) = I_0 \exp{\left(-\left[ (x/R_x)^2 + (z/R_z)^2 \right] \right)}$.

The mass density profile at $t \sim -0.24 \, \mathrm{ps}$ serves as the initial condition for the Particle-In-Cell (PIC) simulation. The conversion is carried out using the \textsc{flash2openpmd} tool \cite{flash2openPMD}. The grid resolution is reduced through cascading interpolation up to six levels of refinement.

The 3D PIC simulations are carried out using the \textsc{picongpu} code version 0.8.0 \cite{PIConGPU}. The dimension of the simulation box is $15 \times 25 \times 15 \, \mathrm{\mu m^3}$. The simulation of the flat target has a resolution of $29 \times 14.6 \times 29 \, \mathrm{nm^3}$, while the nanowire has a resolution of $14.6 \times 14.6 \times 14.6 \, \mathrm{nm^3}$. Each macro-particle species has 8 particle-per-cell, and the shape function used is piecewise quartic spline. To mitigate numerical heating, binomial current interpolation is employed. The target temperatures and charge states obtained from the hydrodynamic simulation vary and are non-uniformly distributed. For simplicity, we assume a uniform temperature and a fully ionized target.

The temporal profile of the main pulse includes the leading edge of the Gaussian pulse from $t \gtrsim -0.24 \, \mathrm{ps}$. The pulse duration is $t_L = 25 \, \mathrm{fs}$ with FWHM focal spot size of $d_\mathrm{FWHM} = 4.2 \, \mathrm{\mu m}$, and wavelength of $\lambda_L = 800 \, \mathrm{nm}$. The laser is p-polarized and focused on the target surface with a peak intensity of $I_0 = 2.8 \times 10^{21} \, \mathrm{W \, cm^{-2}}$, which corresponds to the normalized amplitude of $a_0 = 36$.

The 2D PIC simulations are carried out with reduced physics to track the evolution of electrons and the magnetic field at its position. These simulations have a domain size of $5 \times 5 \, \mathrm{\mu m^2}$, with resolution of $4.8 \times 4.8 \, \mathrm{nm^2}$. The dimension of both targets, and laser parameters remain the same, except that no interaction with laser pre-pulse or rising edge of the main pulse.
\section*{Acknowledgements}
GRK acknowledges major support for this research from the grant “Physics and Astronomy (Project Identification No. RTI4002) Department of Atomic Energy,  Tata Institute of Fundamental Research" and partially from the grant JBR/2021/00039  of the Anusandhan  National Research Foundation (ANRF), both of the Government of India. JFO acknowledges EuroHPC Joint Undertaking for awarding us access to Karolina at IT4Innovations (VŠB-TU), Czechia under project number EHPC-REG-2023R02-006 (DD-23-157); Ministry of Education, Youth and Sports of the Czech Republic through the e-INFRA CZ (ID:90140). These works were partly supported by Contract No. PN23210105 funded by the Romanian Ministry of Research, Innovation and Digitalization and of the Extreme Light Infrastructure Nuclear Physics Phase II, a project co-financed by the Romanian Government and the European Union through the European Regional Development Fund and the Competitiveness Operational Program (Grant No. 1/07.07.2016, COP, ID 1334). Additionally, partial support was given by JSPS Core-to-Core Program, Grant Number JPJSCCA20230003. SI and KAT acknowledges support from IOSIN funds for Facilities of National Interest, and Project $ELI-RO/DFG/2025-013$ IATP-NP 2.0 funded by the Institute of Atomic Physics, Romania. This work was partially supported by JSPS Core-to-Core Program,  (grant number: JPJSCCA20230003) for KAT. Habara acknowledges JSPS KAKENHI Grant Number JP22H01205. We acknowledge ELI-RO/RDI 16 DELPHI, ELI-RO/RDI 28 FLIGHT projects and ELI-RO 21 LGS which supported equipments and materials used in the experiments. We thank Mihail Cernaianu and Petru Ghenuche for encouragement and insights and the entire laser team at ELI-NP for smooth operation of the laser facility.

\section*{Contributions}
A.P., J.F.O. and S.I. contributed equally to this work.  
S.I., in collaboration with D.P. and A.V., fabricated and characterized the nanowire targets. J.F.O. performed the computer simulations. S.D., S.R.,Y.K. participated in the experiment and helped with the post  processing.  A.P. did the data analysis in collaboration with J.F.O. and S.D.. L.T., D.S. and S.I. were involved in the experimental campaign, operation of the diagnostics and data acquisition. R.P. K.S., H.H. provided helpful insights and advice. A.P., J.F.O, S.I. and G.R.K. wrote the  initial draft of the paper. All authors discussed the results, contributed to the interpretation of the data, and reviewed the manuscript. G.R.K., K.A.T.,and P.K.S. conceived and supervised the  entire project.
\par
\begin{center}
\rule{0.5\linewidth}{0.5pt}
\end{center}
\par

\bibliography{ref}

@article{PGibbon_1996,
doi = {10.1088/0741-3335/38/6/001},
url = {https://dx.doi.org/10.1088/0741-3335/38/6/001},
year = {1996},
month = {jun},
publisher = {},
volume = {38},
number = {6},
pages = {769},
author = {P Gibbon and E Förster},
title = {Short-pulse laser - plasma interactions},
journal = {Plasma Physics and Controlled Fusion}
}

@article{10.1063/1.860697,
    author = {Wilks, S. C.},
    title = {Simulations of ultraintense laser–plasma interactions*},
    journal = {Physics of Fluids B: Plasma Physics},
    volume = {5},
    number = {7},
    pages = {2603-2608},
    year = {1993},
    month = {07},
    issn = {0899-8221},
    doi = {10.1063/1.860697},
    url = {https://doi.org/10.1063/1.860697},
    eprint = {https://pubs.aip.org/aip/pfb/article-pdf/5/7/2603/12263345/2603\_1\_online.pdf},
}

@article{electron_nanowire_habara,
    author = {Habara, H. and Honda, S. and Katayama, M. and Sakagami, H. and Nagai, K. and Tanaka, K. A.},
    title = {Efficient energy absorption of intense ps-laser pulse into nanowire target},
    journal = {Physics of Plasmas},
    volume = {23},
    number = {6},
    pages = {063105},
    year = {2016},
    month = {06},
    issn = {1070-664X},
    doi = {10.1063/1.4953092},
    url = {https://doi.org/10.1063/1.4953092},
    eprint = {https://pubs.aip.org/aip/pop/article-pdf/doi/10.1063/1.4953092/15909446/063105_1_online.pdf},
}

@article{preplasma_nanowire,
	author = {Ong, J. F. and Zubarev, A. and Berceanu, A. C. and Cuzminschi, M. and Tesileanu, O.},
	date = {2023/11/24},
	doi = {10.1038/s41598-023-48090-9},
	id = {Ong2023},
	isbn = {2045-2322},
	journal = {Scientific Reports},
	number = {1},
	pages = {20699},
	title = {Nanowire implosion under laser amplified spontaneous emission pedestal irradiation},
	url = {https://doi.org/10.1038/s41598-023-48090-9},
	volume = {13},
	year = {2023}}

@article{parameter_scan_elec_temp,
    author = {Kong, Defeng and Zhang, Guoqiang and Shou, Yinren and Xu, Shirui and Mei, Zhusong and Cao, Zhengxuan and Pan, Zhuo and Wang, Pengjie and Qi, Guijun and Lou, Yao and Ma, Zhiguo and Lan, Haoyang and Wang, Wenzhao and Li, Yunhui and Rubovic, Peter and Veselsky, Martin and Bonasera, Aldo and Zhao, Jiarui and Geng, Yixing and Zhao, Yanying and Fu, Changbo and Luo, Wen and Ma, Yugang and Yan, Xueqing and Ma, Wenjun},
    title = {High-energy-density plasma in femtosecond-laser-irradiated nanowire-array targets for nuclear reactions},
    journal = {Matter and Radiation at Extremes},
    volume = {7},
    number = {6},
    pages = {064403},
    year = {2022},
    month = {11},
    issn = {2468-2047},
    doi = {10.1063/5.0120845},
    url = {https://doi.org/10.1063/5.0120845},
    eprint = {https://pubs.aip.org/aip/mre/article-pdf/doi/10.1063/5.0120845/16562833/064403_1_online.pdf},
}

@article{nat_photonics_purvis2013relativistic,
  title={Relativistic plasma nanophotonics for ultrahigh energy density physics},
  author={Purvis, Michael A and Shlyaptsev, Vyacheslav N and Hollinger, Reed and Bargsten, Clayton and Pukhov, Alexander and Prieto, Amy and Wang, Yong and Luther, Bradley M and Yin, Liang and Wang, Shoujun and others},
  journal={Nature photonics},
  volume={7},
  number={10},
  pages={796--800},
  year={2013},
  publisher={Nature Publishing Group UK London}
}

@article{prl_kaymak2016nanoscale,
  title={Nanoscale ultradense Z-pinch formation from laser-irradiated nanowire arrays},
  author={Kaymak, Vural and Pukhov, Alexander and Shlyaptsev, Vyacheslav N and Rocca, Jorge J},
  journal={Physical review letters},
  volume={117},
  number={3},
  pages={035004},
  year={2016},
  publisher={APS}
}

@article{nat_photonics_hollinger2020extreme,
  title={Extreme ionization of heavy atoms in solid-density plasmas by relativistic second-harmonic laser pulses},
  author={Hollinger, R and Wang, S and Wang, Y and Moreau, A and Capeluto, Maria Gabriela and Song, H and Rockwood, A and Bayarsaikhan, E and Kaymak, V and Pukhov, A and others},
  journal={Nature Photonics},
  volume={14},
  number={10},
  pages={607--611},
  year={2020},
  publisher={Nature Publishing Group UK London}
}

@article{bargsten2017energy,
  title={Energy penetration into arrays of aligned nanowires irradiated with relativistic intensities: Scaling to terabar pressures},
  author={Bargsten, Clayton and Hollinger, Reed and Capeluto, Maria Gabriela and Kaymak, Vural and Pukhov, Alexander and Wang, Shoujun and Rockwood, Alex and Wang, Yong and Keiss, David and Tommasini, Riccardo and others},
  journal={Science advances},
  volume={3},
  number={1},
  pages={e1601558},
  year={2017},
  publisher={American Association for the Advancement of Science}
}

@article{prx_samsonova2019relativistic,
  title={Relativistic interaction of long-wavelength ultrashort laser pulses with nanowires},
  author={Samsonova, Zhanna and H{\"o}fer, Sebastian and Kaymak, Vural and Ali{\v{s}}auskas, Skirmantas and Shumakova, Valentina and Pug{\v{z}}lys, Audrius and Baltu{\v{s}}ka, Andrius and Siefke, Thomas and Kroker, Stefanie and Pukhov, Alexander and others},
  journal={Physical Review X},
  volume={9},
  number={2},
  pages={021029},
  year={2019},
  publisher={APS}
}

@article{pop_eftekhari2022laser,
  title={Laser energy absorption and x-ray generation in nanowire arrays irradiated by relativistically intense ultra-high contrast femtosecond laser pulses},
  author={Eftekhari-Zadeh, E and Bl{\"u}mcke, MS and Samsonova, Z and Loetzsch, R and Uschmann, I and Zapf, M and Ronning, C and Rosmej, ON and Kartashov, D and Spielmann, C},
  journal={Physics of Plasmas},
  volume={29},
  number={1},
  year={2022},
  publisher={AIP Publishing}
}

@article{prl_jiang2016microengineering,
  title={Microengineering laser plasma interactions at relativistic intensities},
  author={Jiang, S and Ji, LL and Audesirk, H and George, KM and Snyder, J and Krygier, A and Poole, P and Willis, C and Daskalova, R and Chowdhury, E and others},
  journal={Physical review letters},
  volume={116},
  number={8},
  pages={085002},
  year={2016},
  publisher={APS}
}

@article{scientific_reports_fedeli2018ultra,
  title={Ultra-intense laser interaction with nanostructured near-critical plasmas},
  author={Fedeli, Luca and Formenti, Arianna and Cialfi, Lorenzo and Pazzaglia, Andrea and Passoni, Matteo},
  journal={Scientific reports},
  volume={8},
  number={1},
  pages={3834},
  year={2018},
  publisher={Nature Publishing Group UK London}
}

@article{pop_park2021absolute,
  title={Absolute laser energy absorption measurement of relativistic 0.7 ps laser pulses in nanowire arrays},
  author={Park, Jaebum and Tommasini, Riccardo and Shepherd, R and London, RA and Bargsten, Clayton and Hollinger, Reed and Capeluto, Maria Gabriela and Shlyaptsev, VN and Hill, MP and Kaymak, V and others},
  journal={Physics of Plasmas},
  volume={28},
  number={2},
  year={2021},
  publisher={AIP Publishing}
}

@article{prr_ong2021electron,
  title={Electron transport in a nanowire irradiated by an intense laser pulse},
  author={Ong, JF and Ghenuche, P and Tanaka, KA},
  journal={Physical Review Research},
  volume={3},
  number={3},
  pages={033262},
  year={2021},
  publisher={APS}
}

@article{apl_ji2010efficient,
  title={Efficient generation and transportation of energetic electrons in a carbon nanotube array target},
  author={Ji, Yanling and Jiang, Gang and Wu, Weidong and Wang, Chaoyang and Gu, Yuqiu and Tang, Yongjian},
  journal={Applied Physics Letters},
  volume={96},
  number={4},
  year={2010},
  publisher={AIP Publishing}
}

@article{pop_zhao2010acceleration,
  title={Acceleration and guiding of fast electrons by a nanobrush target},
  author={Zhao, Zongqing and Cao, Lihua and Cao, Leifeng and Wang, Jian and Huang, Wenzhong and Jiang, Wei and He, Yingling and Wu, Yuchi and Zhu, Bin and Dong, Kegong and others},
  journal={Physics of Plasmas},
  volume={17},
  number={12},
  year={2010},
  publisher={AIP Publishing}
}

@article{pop_cao2010enhanced,
  title={Enhanced absorption of intense short-pulse laser light by subwavelength nanolayered target},
  author={Cao, Lihua and Gu, Yuqiu and Zhao, Zongqing and Cao, Leifeng and Huang, Wenzhong and Zhou, Weimin and He, XT and Yu, Wei and Yu, MY},
  journal={Physics of Plasmas},
  volume={17},
  number={4},
  year={2010},
  publisher={AIP Publishing}
}

@article{laserphotonics_ankit,
author = {Dulat, Ankit and Rakeeb, Sk and Dam, Sagar and Lad, Amit D. and Ved, Yash M. and Kruk, Sergey and Kumar, G. Ravindra},
title = {Coherent Control of Relativistic Electron Dynamics in Plasma Nanophotonics},
journal = {Laser \& Photonics Reviews},
volume = {19},
number = {5},
pages = {2401570},
doi = {https://doi.org/10.1002/lpor.202401570},
url = {https://onlinelibrary.wiley.com/doi/abs/10.1002/lpor.202401570},
eprint = {https://onlinelibrary.wiley.com/doi/pdf/10.1002/lpor.202401570},
year = {2025}
}

@article{Rocca:24,
author = {Jorge J. Rocca and Maria G. Capeluto and Reed C. Hollinger and Shoujun Wang and Yong Wang and G. Ravindra Kumar and Amit D. Lad and Alexander Pukhov and Vyacheslav N. Shlyaptsev},
journal = {Optica},
number = {3},
pages = {437--453},
publisher = {Optica Publishing Group},
title = {Ultra-intense femtosecond laser interactions with aligned nanostructures},
volume = {11},
month = {Mar},
year = {2024},
url = {https://opg.optica.org/optica/abstract.cfm?URI=optica-11-3-437},
doi = {10.1364/OPTICA.510542},
}

@article{PRR_aprajit_2025,
  title = {Role of femtosecond prestructure of intense harmonic pulses in relativistic laser-solid interactions},
  author = {Aparajit, C. and Choudhary, Anandam and Dulat, Ankit and Grech, Mickael and Marini, Samuel and Lad, Amit D. and Ved, Yash M. and Raynaud, Mich\`ele and Riconda, Caterina and Kumar, G. Ravindra},
  journal = {Phys. Rev. Res.},
  volume = {7},
  issue = {3},
  pages = {L032063},
  numpages = {7},
  year = {2025},
  month = {Sep},
  publisher = {American Physical Society},
  doi = {10.1103/PhysRevResearch.7.L032063},
  url = {https://link.aps.org/doi/10.1103/PhysRevResearch.7.L032063}
}

@article{Sci-rep_amit_lad,
	author = {Lad, Amit D. and Mishima, Y. and Singh, Prashant Kumar and Li, Boyuan and Adak, Amitava and Chatterjee, Gourab and Brijesh, P. and Dalui, Malay and Inoue, M. and Jha, J. and Tata, Sheroy and Trivikram, M. and Krishnamurthy, M. and Chen, Min and Sheng, Z. M. and Tanaka, K. A. and Kumar, G. Ravindra and Habara, H.},
	date = {2022/10/07},
	doi = {10.1038/s41598-022-21210-7},
	id = {Lad2022},
	isbn = {2045-2322},
	journal = {Scientific Reports},
	number = {1},
	pages = {16818},
	title = {Luminous, relativistic, directional electron bunches from an intense laser driven grating plasma},
	url = {https://doi.org/10.1038/s41598-022-21210-7},
	volume = {12},
	year = {2022}}

@ARTICLE{Gheorghiu2021,
       author = {{Gheorghiu}, Cristina C. and {Ionescu}, Stefania C. and {Ghenuche}, Petru and {Cernaianu}, Mihail O. and {Doria}, Domenico and {Popa}, Daniel and {Leca}, Victor},
        title = "{Structuring free-standing foils for laser-driven particle acceleration experiments}",
      journal = {Frontiers in Physics},
         year = 2021,
        month = 9,
       volume = {9},
          eid = {515},
        pages = {515},
          doi = {10.3389/fphy.2021.727498},
       }

@article{Masuda1995,
author = {Hideki Masuda  and Kenji Fukuda },
title = {Ordered Metal Nanohole Arrays Made by a Two-Step Replication of Honeycomb Structures of Anodic Alumina},
journal = {Science},
volume = {268},
number = {5216},
pages = {1466-1468},
year = {1995},
doi = {10.1126/science.268.5216.1466},
}

@ARTICLE{PIConGPU, 
author={H. {Burau} and R. {Widera} and W. {Hönig} and G. {Juckeland} and A. {Debus} and T. {Kluge} and U. {Schramm} and T. E. {Cowan} and R. {Sauerbrey} and M. {Bussmann}}, 
journal={IEEE Transactions on Plasma Science}, 
title={PIConGPU: A Fully Relativistic Particle-in-Cell Code for a GPU Cluster}, 
year={2010}, 
volume={38}, 
number={10}, 
pages={2831-2839}, 
doi={10.1109/TPS.2010.2064310}, 
ISSN={1939-9375}, 
month={Oct},}

@article{flash,
  title = {FLASH: An adaptive mesh hydrodynamics code for modeling astrophysical
          thermonuclear flashes},
  author = {Fryxell, Bruce and Olson, Kevin and Ricker, Paul and Timmes, FX and
            Zingale, Michael and Lamb, DQ and MacNeice, Peter and Rosner, Robert
            and Truran, JW and Tufo, H},
  journal = {The Astrophysical Journal Supplement Series},
  volume = {131},
  number = {1},
  pages = {273},
  year = {2000},
  publisher = {IOP Publishing},
  doi = {https://doi.org/10.1086/317361},
  url = {https://dx.doi.org/10.1086/317361},
}

@misc{flash2openPMD,
  author = {J. F. Ong},
  howpublished = {https://github.com/ELI-NP/flash2openPMD.git},
  title = {flash2openPMD},
  year = {2022}
}

@book{fortov_Extreme_states_of_matter,
  title={Extreme states of matter: high energy density physics},
  author={Fortov, Vladimir E},
  volume={216},
  year={2015},
  publisher={Springer}
}

@book{Kruer, 
   AUTHOR={W.L.Kruer}, 
   TITLE={The physics of laser plasma interaction}, 
   PUBLISHER={Westview Press},
   ADDRESS={Boulder, Colorado},
   YEAR=2003 
 }

@article{ankit2022spectralinterferometry,
  title={Subpicosecond pre-plasma dynamics of a high contrast, ultraintense laser--solid target interaction},
  author={Dulat, Ankit and Aparajit, C and Choudhary, Anandam and Lad, Amit D and Ved, Yash M and Paradkar, BS and Ravindra Kumar, G},
  journal={Optics Letters},
  volume={47},
  number={21},
  pages={5684--5687},
  year={2022},
  publisher={Optica Publishing Group},
doi={https://doi.org/10.1364/OL.461452}
}

@article{ciappina2017attosecond,
  title={Attosecond physics at the nanoscale},
  author={Ciappina, Marcello F and P{\'e}rez-Hern{\'a}ndez, Jos{\'e} A and Landsman, Alexandra S and Okell, William A and Zherebtsov, Sergey and F{\"o}rg, Benjamin and Sch{\"o}tz, Johannes and Seiffert, Lennart and Fennel, Thomas and Shaaran, Tahir and others},
  journal={Reports on Progress in Physics},
  volume={80},
  number={5},
  pages={054401},
  year={2017},
  publisher={IOP Publishing}
}

@article{moreau2019enhanced,
  title={Enhanced electron acceleration in aligned nanowire arrays irradiated at highly relativistic intensities},
  author={Moreau, A and Hollinger, R and Calvi, C and Wang, S and Wang, Y and Capeluto, Maria Gabriela and Rockwood, A and Curtis, A and Kasdorf, S and Shlyaptsev, VN and others},
  journal={Plasma Physics and Controlled Fusion},
  volume={62},
  number={1},
  pages={014013},
  year={2019},
  publisher={IOP Publishing}
}

@article{mourou_Tajima_2006optics_RMP,
  title={Optics in the relativistic regime},
  author={Mourou, Gerard A and Tajima, Toshiki and Bulanov, Sergei V},
  journal={Reviews of modern physics},
  volume={78},
  doi={https://doi.org/10.1103/RevModPhys.78.309},
  number={2},
  pages={309--371},
  year={2006},
  publisher={APS}
}

@article{kiriyama2012temporal,
  title={Temporal contrast enhancement of petawatt-class laser pulses},
  author={Kiriyama, Hiromitsu and Shimomura, Takuya and Sasao, Hajime and Nakai, Yoshiki and Tanoue, Manabu and Kondo, Shuji and Kanazawa, Shuhei and Pirozhkov, Alexander S and Mori, Michiaki and Fukuda, Yuji and others},
  journal={Optics Letters},
  volume={37},
  doi={https://doi.org/10.1364/OL.37.003363},
  number={16},
  pages={3363--3365},
  year={2012},
  publisher={Optical Society of America}
}

@article{choi2020highly,
  title={Highly efficient double plasma mirror producing ultrahigh-contrast multi-petawatt laser pulses},
  author={Choi, Il Woo and Jeon, Cheonha and Lee, Seong Geun and Kim, Seung Yeon and Kim, Tae Yun and Kim, I Jong and Lee, Hwang Woon and Woo Yoon, Jin and Sung, Jae Hee and Lee, Seong Ku and others},
  journal={Optics Letters},
  volume={45},
  doi = {https://doi.org/10.1364/OL.409749},
  number={23},
  pages={6342--6345},
  year={2020},
  publisher={Optical Society of America}
}

@article{contrast1998itatani,
  title={Suppression of the amplified spontaneous emission in chirped-pulse-amplification lasers by clean high-energy seed-pulse injection},
  author={Itatani, J and Faure, J and Nantel, M and Mourou, G and Watanabe, SJOC},
  journal={Optics Communications},
  volume={148},
  number={1-3},
  pages={70--74},
  year={1998},
  publisher={Elsevier},
doi={https://doi.org/10.1016/S0030-4018(97)00638-X}
}

@article{plasmamirror2006wittmanntowards,
  title={Towards ultrahigh-contrast ultraintense laser pulses—complete characterization of a double plasma-mirror pulse cleaner},
  author={Wittmann, T and Geindre, Jean-Paul and Audebert, P and Marjoribanks, RS and Rousseau, Jean-Philippe and Burgy, Fr{\'e}d{\'e}ric and Douillet, D and Lefrou, T and Phuoc, K Ta and Chambaret, Jean-Paul},
  journal={Review of scientific instruments},
  volume={77},
  number={8},
  year={2006},
  publisher={AIP Publishing},
doi={https://doi.org/10.1063/1.2234850}
}

@inproceedings{plasmamirror2018foldes,
  title={Plasma Mirrors for Cleaning Laser Pulses from the Infrared to the Ultraviolet},
  author={F{\"o}ldes, Istv{\'a}n B and Gilicze, Barnab{\'a}s and Kov{\'a}cs, Zsolt and Szatm{\'a}ri, S{\'a}ndor},
  booktitle={EPJ Web of Conferences},
  volume={167},
  pages={04001},
  year={2018},
  organization={EDP Sciences},
doi={https://doi.org/10.1051/epjconf/201816704001}
}

@article{Ionescu2025,
  title={Highly ordered vertical nickel nanotubes and nanowires on thin substrate for high power lasers experiments},
  author={Ionescu, Stefania C and Gheorghiu, Cristina C and Lupu, Valentin and Zai, Maria-Iulia and Magureanu, Alexandru and Dreghici, Dragana B and McCay, Adrian and Molloy, Daniel and Ahmed, Hamad and Borghesi, Marco and others},
  journal={Discover Nano},
  doi={https://doi.org/10.1186/s11671-025-04394-5},
  volume={20},
  number={1},
  pages={219},
  year={2025},
  publisher={Springer}
}

@article{Wong2024,
    author = {Wong, C.-S. and Strehlow, J. and Broughton, D. P. and Luedtke, S. V. and Huang, C.-K. and Bogale, A. and Fitzgarrald, R. and Nedbailo, R. and Schmidt, J. L. and Schmidt, T. R. and Twardowski, J. and Van Pelt, A. and Alvarez, M. Alvarado and Junghans, A. and Mix, L. T. and Reinovsky, R. E. and Rusby, D. R. and Wang, Z. and Wolfe, B. and Albright, B. J. and Batha, S. H. and Palaniyappan, S.},
    title = {Robust unfolding of MeV x-ray spectra from filter stack spectrometer data},
    journal = {Review of Scientific Instruments},
    volume = {95},
    number = {2},
    pages = {023301},
    year = {2024},
    month = {02},
    issn = {0034-6748},
    doi = {10.1063/5.0190679},
}

@article{commissioning_1PW,
    author = {Cernaianu, M. O. and Ghenuche, P. and Rotaru, F. and Tudor, L. and Chalus, O. and Gheorghiu, C. and Popescu, D. C. and Gugiu, M. and Balascuta, S. and Magureanu, A. and Tataru, M. and Horny, V. and Corobean, B. and Dancus, I. and Alincutei, A. and Asavei, T. and Diaconescu, B. and Dinca, L. and Dreghici, D. B. and Ghita, D. G. and Jalba, C. and Leca, V. and Lupu, A. M. and Nastasa, V. and Negoita, F. and Patrascoiu, M. and Schimbeschi, F. and Stutman, D. and Ticos, C. and Ursescu, D. and Arefiev, A. and Tomassini, P. and Malka, V. and Gales, S. and Tanaka, K. A. and Ur, C. A. and Doria, D.},
    title = {Commissioning of the 1 PW experimental area at ELI-NP using a short focal parabolic mirror for proton acceleration},
    journal = {Matter and Radiation at Extremes},
    volume = {10},
    number = {2},
    pages = {027204},
    year = {2025},
    month = {03},
    issn = {2468-2047},
    doi = {10.1063/5.0241077},
    url = {https://doi.org/10.1063/5.0241077},
    eprint = {https://pubs.aip.org/aip/mre/article-pdf/doi/10.1063/5.0241077/20439722/027204_1_5.0241077.pdf},
}

@article{commissoning_1PW_2,
    author = {Tanaka, K. A. and Spohr, K. M. and Balabanski, D. L. and Balascuta, S. and Capponi, L. and Cernaianu, M. O. and Cuciuc, M. and Cucoanes, A. and Dancus, I. and Dhal, A. and Diaconescu, B. and Doria, D. and Ghenuche, P. and Ghita, D. G. and Kisyov, S. and Nastasa, V. and Ong, J. F. and Rotaru, F. and Sangwan, D. and Söderström, P.-A. and Stutman, D. and Suliman, G. and Tesileanu, O. and Tudor, L. and Tsoneva, N. and Ur, C. A. and Ursescu, D. and Zamfir, N. V.},
    title = {Current status and highlights of the ELI-NP research program},
    journal = {Matter and Radiation at Extremes},
    volume = {5},
    number = {2},
    pages = {024402},
    year = {2020},
    month = {03},
    issn = {2468-2047},
    doi = {10.1063/1.5093535},
    url = {https://doi.org/10.1063/1.5093535},
    eprint = {https://pubs.aip.org/aip/mre/article-pdf/doi/10.1063/1.5093535/15750185/024402_1_online.pdf},
}

@article{PP_rajeev_PRL_2003,
  title = {Metal Nanoplasmas as Bright Sources of Hard X-Ray Pulses},
  author = {Rajeev, P. P. and Taneja, P. and Ayyub, P. and Sandhu, A. S. and Kumar, G. Ravindra},
  journal = {Phys. Rev. Lett.},
  volume = {90},
  issue = {11},
  pages = {115002},
  numpages = {4},
  year = {2003},
  month = {Mar},
  publisher = {American Physical Society},
  doi = {10.1103/PhysRevLett.90.115002},
  url = {https://link.aps.org/doi/10.1103/PhysRevLett.90.115002}
}

@article{S_Kahalay_PRL_2008,
  title = {Near-Complete Absorption of Intense, Ultrashort Laser Light by Sub-$\ensuremath{\lambda}$ Gratings},
  author = {Kahaly, Subhendu and Yadav, S. K. and Wang, W. M. and Sengupta, S. and Sheng, Z. M. and Das, A. and Kaw, P. K. and Kumar, G. Ravindra},
  journal = {Phys. Rev. Lett.},
  volume = {101},
  issue = {14},
  pages = {145001},
  numpages = {4},
  year = {2008},
  month = {Sep},
  publisher = {American Physical Society},
  doi = {10.1103/PhysRevLett.101.145001},
  url = {https://link.aps.org/doi/10.1103/PhysRevLett.101.145001}
}
\bibliographystyle{ieeetr}
\appendix
\end{document}